\documentclass[twoside,twocolumn,english,prl]{revtex4-1}

\usepackage[T1]{fontenc}
\usepackage[latin9]{inputenc}
\usepackage{amsmath}
\usepackage{graphicx}

\makeatletter
\@ifundefined{date}{}{\date{}}
\makeatother

\usepackage{babel}
\begin{document}

\title{Role of Hund's splitting in electronic phase competition in Pb$_{1-x}$Sn$_{x}$Te}

\author{S.Kundu and V.Tripathi}

\affiliation{Department of Theoretical Physics,Tata Institute of Fundamental Research,Homi
Bhabha Road, Navy Nagar, Colaba, Mumbai-400005}

\date{\today}
\begin{abstract}
We study the effect of Hund's splitting of repulsive interactions
on electronic phase transitions in the multiorbital topological crystalline
insulator Pb$_{1-x}$Sn$_{x}$Te, when the chemical potential is tuned
to the vicinity of low-lying Type-II Van Hove singularities. Nontrivial
Berry phases associated with the Bloch states impart momentum-dependence
to electron interactions in the relevant band. We use a multipatch
parquet renormalization group (RG) analysis for studying the competition
of different electronic phases, and find that if the dominant fixed-point
interactions correspond to antiparallel spin configurations, then
a chiral $p$-wave Fulde-Ferrell-Larkin-Ovchinnikov(FFLO) state is
favored, otherwise, none of the commonly encountered electronic instabilities
occur within the one-loop parquet RG approach.
\end{abstract}
\maketitle
Topological crystalline insulators (TCIs) have low-energy surface
states in certain high symmetry directions, protected by crystalline
symmetry \cite{fu2011topological}. Unlike conventional Z$_{2}$ topological
insulators \cite{hasan2010colloquium,konig2008quantum,moore2010birth,qi2011topological},
the nature of these low-energy states is sensitive to the surface
orientation. In particular, it has been shown in the recently discovered
TCI Pb$_{1-x}$Sn$_{x}$Te \cite{dziawa2012topological,hsieh2012topological,tanaka2012experimental,xu2012observation}
that the band structure of the (001) surface allows for the presence
of Type-II Van Hove singularities \cite{PhysRevB.92.035132}, with
a diverging density of states, which opens up the possibility of a
variety of competing Fermi-surface instabilities brought about by
weak repulsive interparticle interactions \cite{dzyaloshinskii1987maximal,schulz1987superconductivity,lederer1987antiferromagnetism,PhysRevLett.112.070403,PhysRevB.89.144501}.
In particular, the parquet approximation\textbf{ }for studying competing
phases in a system with multiple Fermi pockets has proved very useful
in the context of unconventional superconductivity \cite{mineev1999introduction,norman2011challenge,sigrist1991phenomenological}
in cuprates \cite{furukawa1998truncation}, graphene \cite{nandkishore2012chiral}
and semimetal thin films \cite{PhysRevB.93.155108}. However, in a
multiorbital system like Pb$_{1-x}$Sn$_{x}$Te, phase competition
needs to be studied taking into account the effect of Hund's splitting
of interactions. The importance of Hund's coupling has generally been
underemphasized in parquet renormalization group analyses of multiorbital
systems for reasons of convenience, but recent developments show that
Hund's coupling may play an important role in electronic instabilities
of multiorbital systems\textbf{ }\cite{yuan2015triplet,vafek2017hund}. 

In this paper, we employ a multipatch parquet renormalization group
(RG) analysis including Hund's splitting effects, and show that even
relatively small amounts of Hund's splitting can have a dramatic effect
on the very existence of electronic instabilities on the surface of
Pb$_{1-x}$Sn$_{x}$Te.  Depending on the sign of the Hund's splitting,
we find that away from perfect nesting, either a chiral $p$-wave
FFLO \cite{PhysRev.135.A550,larkin1964nonuniform} state is stabilized
or none of the commonly encountered electronic instabilities occur
at the level of the one-loop parquet approach. A characteristic feature
of Pb$_{1-x}$Sn$_{x}$Te is that the surface bands are effectively
spinless, which rules out $s$-wave pairing, that would otherwise
prevail over $p$-wave pairing in the presence of nonmagnetic disorder
\cite{PhysRev.131.1553,RevModPhys.75.657,anderson1959theory}. 

The topological crystalline insulator surface that we consider offers
certain natural advantages from an experimental point of view. It
provides two-dimensional Van Hove singularities which are accessible
through a small change in doping, unlike, say, graphene, where a very
high level of doping is required.\textbf{ }Interestingly, as we show
below, the $p$-wave symmetry originates not from intrinsic Fermi
surface deformations, but from the nontrivial Berry phases associated
with the topological states. This is reminiscent of chiral $p$-wave
superconductivity enabled by a topological Berry phase in fermionic
cold atom systems with attractive momentum-independent interactions
\cite{zhang2008p}. We argue that the $p$-wave superconductivity
on the TCI surface is more robust against potential disorder \cite{PhysRevLett.109.187003,nagai2015robust}
than in, say, Sr$_{2}$RuO$_{4}$ \cite{PhysRevLett.80.161}. Moreover,
the $p$-wave superconductivity here is intrinsic, unlike proximity-induced
$p$-wave superconductivity on topological insulator surfaces where
recently Majorana fermions have been detected \cite{he2017chiral}.
Finally, such an FFLO state in a pure solid state system in the absence
of an applied magnetic field is a rather unusual occurrence (see,
e.g. Refs \cite{200815637} and \cite{hsu2017topological}). Ref.
\cite{cho2012superconductivity} also discusses an intranode FFLO
pairing in a doped Weyl semimetal, although the stability of such
a state in this system is still a controversial issue \cite{bednik2015superconductivity,wei2014odd,zhou2016superconductivity}. 

\begin{figure}
\includegraphics[width=1.0\columnwidth]{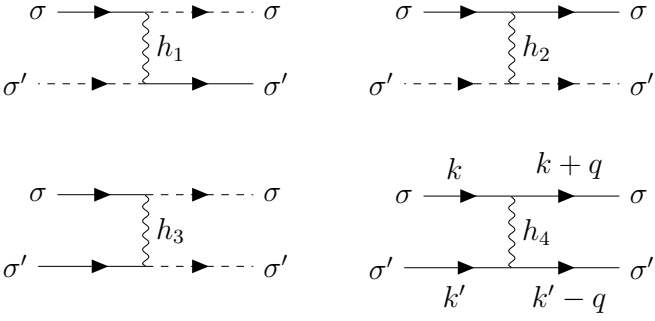}

\caption{\label{fig:The-RG-couplings}The different types of Coulomb interaction
processes in our low-energy model (Eq.\ref{eq:3}). The solid lines
and dashed lines denote two different patches $\overline{X_{1}}$
and $\overline{X_{2}}$ in momentum space, on the (001) surface. All
the vertices have momentum-dependences as indicated for $h_{4}$.
The $\sigma$'s refer to the particular spin components of the (spinor)
wavefunctions associated with the bands under consideration (see text
for more details). }
 
\end{figure}

The band gap minima of IV-VI semiconductors are located at the four
$L$ points in the FCC Brillouin zone.\textbf{ }In \cite{liu2013two},
the TCI surface states are classified into two types: \emph{Type-I},
for which all four $L$-points are projected to the different time-reversal
invariant momenta(TRIM) in the surface Brillouin zone, and \emph{Type-II},
for which different $L$-points are projected to the same surface
momentum. The (001) surface falls into the latter class of surfaces,
for which the $L_{1}$ and $L_{2}$ points are projected to the $\overline{X_{1}}$
point on the surface, and the $L_{3}$ and $L_{4}$ points are projected
to the symmetry-related $\overline{X_{2}}$ point. This leads to two
coexisting massless Dirac fermions at $\overline{X_{1}}$ arising
from the $L_{1}$ and the $L_{2}$ valley, respectively, and likewise
at $\overline{X_{2}}$. The k.p Hamiltonian close to the point $\overline{X_{1}}$
on the (001) surface is derived on the basis of a symmetry analysis
in \cite{liu2013two}, and is given by 
\begin{equation}
H_{\overline{X_{1}}}(k)=(v_{x}k_{x}s_{y}-v_{y}k_{y}s_{x})+m\tau_{x}+\delta s_{x}\tau_{y},\label{eq:1}
\end{equation}
where $k$ is measured with respect to $\overline{X_{1}}$, $\overrightarrow{s}$
is a set of Pauli matrices associated with the two spin components
associated with each valley, $\tau$ operates in valley space, and
the terms $m$ and $\delta$, which are off-diagonal in valley space,
are added to describe intervalley scattering. The band dispersion
and constant energy contours for the above surface Hamiltonian undergo
a Lifshitz transition with increasing energy away from the Dirac point,
and when the Fermi surface is at $\delta=26$ meV (as taken from \cite{liu2013two})
two saddle points $\overline{S_{1}}$ and $\overline{S_{2}}$ at momenta
$(\pm\frac{m}{v_{x}},0)$ lead to a Van-Hove singularity in the density
of states. A similar situation arises at the point $\overline{X_{2}}$. 

In addition to the noninteracting part of the Hamiltonian described
in Eq.1 above, we now consider interactions between surface
electrons corresponding to different valleys and spins, which gives
rise to the following terms in the Hamiltonian-
\begin{equation}
H_{I}=\frac{1}{2}\sum_{a,b,c,d,\sigma,\sigma^{\prime}}U_{abcd}^{\sigma\sigma^{\prime}}c_{\sigma a}^{\dagger}c_{\sigma^{\prime}b}^{\dagger}c_{\sigma^{\prime}c}c_{\sigma d}\label{eq:2}
\end{equation}
where $a,b,c,d$ refer to different valleys (which are either all
the same, same in pairs or all different in the above sum) and $\sigma,\sigma^{\prime}$
refer to spins. Here, we consider $U_{abcd}^{\sigma\sigma^{\prime}}=U_{1}^{\sigma\sigma^{\prime}}$when
$(a,c)$ belong to one $\overline{X}$-point (i.e. the L-valleys corresponding
to $(a,c)$ are projected to one of the $\overline{X}$-points) and
$(b,d)$ belong to the other $\overline{X}$-point. Similarly, $U_{abcd}^{\sigma\sigma^{\prime}}=U_{2}^{\sigma\sigma^{\prime}}$
when $(b,c)$ belong to one $\overline{X}$-point and $(a,d)$ belong
to the other, $U_{3}^{\sigma\sigma^{\prime}}$ when $(a,b)$ belong
to one $\overline{X}$-point and $(c,d)$ to the other, and $U_{4}^{\sigma\sigma^{\prime}}$when
$a$,$b$,$c$ and $d$ all correspond to L-points projected to the
same $\overline{X}$-point. The interactions depend only on the relative
orientations of the spins, for example, $U^{\sigma\sigma^{\prime}}$
can be written as $U^{\sigma\sigma}\delta_{\sigma\sigma^{\prime}}+U^{\sigma\overline{\sigma}}(1-\delta_{\sigma\sigma^{\prime}})$.
In our analysis, we have projected the interactions between electrons
in the valley-spin picture to the positive-energy band lying closest
to the Van-Hove singularities \cite{seesi}. The resulting multiplicative
form factors $u_{\sigma ai}$(for a transformation from valley $a$,
spin $\sigma$ to the $i$th band) lend a momentum dependence to the
effective pairing interactions obtained upon projection. We find that
the spin $\uparrow$ components of the form factors have an $\exp[i\theta_{k}]$
dependence in momentum space and transform as $\ell=1$ objects, whereas
the phase of the spin $\downarrow$ components remains unchanged upon
advancing by an angle of $2\pi$ around the $\overline{X_{r}}$($r=1,2$)
points, and these show an $\ell=0$ angular dependence. These additional
phase factors arise from the Berry phases associated with the surface
states of the crystalline topological insulator. After projecting
to the two bands intersecting with the Fermi level, we obtain the
following low-energy theory
\begin{align}
L & =\sum_{i}\psi_{i}^{\dagger}(\partial_{\tau}-\epsilon_{k}+\mu)\psi_{i}-\sum_{i,\sigma,\sigma^{\prime}}\frac{1}{2}h_{4}^{\sigma\sigma^{\prime}}\psi_{i}^{\dagger}\psi_{i}^{\dagger}\psi_{i}\psi_{i}\nonumber \\
 & \qquad-\sum_{i\neq j,\sigma,\sigma^{\prime}}\frac{1}{2}(h_{1}^{\sigma\sigma^{\prime}}\psi_{i}^{\dagger}\psi_{j}^{\dagger}\psi_{i}\psi_{j}+h_{2}^{\sigma\sigma^{\prime}}\psi_{i}^{\dagger}\psi_{j}^{\dagger}\psi_{j}\psi_{i}\nonumber \\
 & \qquad+h_{3}^{\sigma\sigma^{\prime}}\psi_{i}^{\dagger}\psi_{i}^{\dagger}\psi_{j}\psi_{j})\nonumber \\
 & \qquad=\sum_{i}\psi_{i}^{\dagger}(\partial_{\tau}-\epsilon_{k}+\mu)\psi_{i}-(h_{4}^{0}+h_{4}^{1})\psi_{i}^{\dagger}\psi_{i}^{\dagger}\psi_{i}\psi_{i}\nonumber \\
 & \qquad-\sum_{i\neq j}((h_{1}^{0}+h_{1}^{1})\psi_{i}^{\dagger}\psi_{j}^{\dagger}\psi_{i}\psi_{j}+(h_{2}^{0}+h_{2}^{1})\psi_{i}^{\dagger}\psi_{j}^{\dagger}\psi_{j}\psi_{i}\nonumber \\
 & \qquad+(h_{3}^{0}+h_{3}^{1})\psi_{i}^{\dagger}\psi_{i}^{\dagger}\psi_{j}\psi_{j})\label{eq:3}
\end{align}
 with $h_{r}^{0}=\frac{1}{2}\sum_{\sigma}h_{r}^{\sigma\sigma}$ and
$h_{r}^{1}=\frac{1}{2}\sum_{\sigma}h_{r}^{\sigma\overline{\sigma}}$
where the quadratic noninteracting part comes from the model in Eq.1.
The chemical potential value $\mu$=0 corresponds to the system being
doped to the Van Hove singularities. Here $h_{4}$ refers to different
scattering processes within a band $i$, whereas $h_{1}$,$h_{2}$
and $h_{3}$ refer to exchange processes, Coulomb interactions and
pair hopping between electrons corresponding to the two different
bands under consideration (see Fig. \ref{fig:The-RG-couplings}).
Due to the distinct phase dependences associated with the form factors
corresponding to spins $\uparrow$ and $\downarrow$, the effective
interactions $h_{r}$ after projection to the low-energy bands also
either have a phase factor of $\exp[i(\theta_{k}-\theta_{k^{\prime}})]$
(for spin-antiparallel configurations) and behave as $\ell=1$ objects,
or have no additional phase factors (for spin-parallel configurations)
and behave as $\ell=0$ objects. The coupling constants $h_{r}^{0}\propto h_{r}^{\sigma\sigma}$
and $h_{r}^{1}\propto h_{r}^{\sigma\overline{\sigma}}$ respectively
correspond to $\ell=0$ and $\ell=1$ angular momentum components
of the interaction in our simplified model in Eq.\ref{eq:3} above.
It is important to note that although the surface bands are effectively
spinless, we associate spin indices $\sigma\sigma^{\prime}$(or equivalently
the superscripts $0$ and $1$) with the interactions $h_{r}$ in
the different scattering channels $r$, due to the phase dependences
associated with interactions between electrons with different spin
configurations. In doing so, we allow for the Coulomb interactions
between electrons to depend on the spin configuration being considered,
thereby incorporating the effects of Hund's splitting of interactions
in our treatment. 

\begin{figure}
\includegraphics[width=1.0\columnwidth]{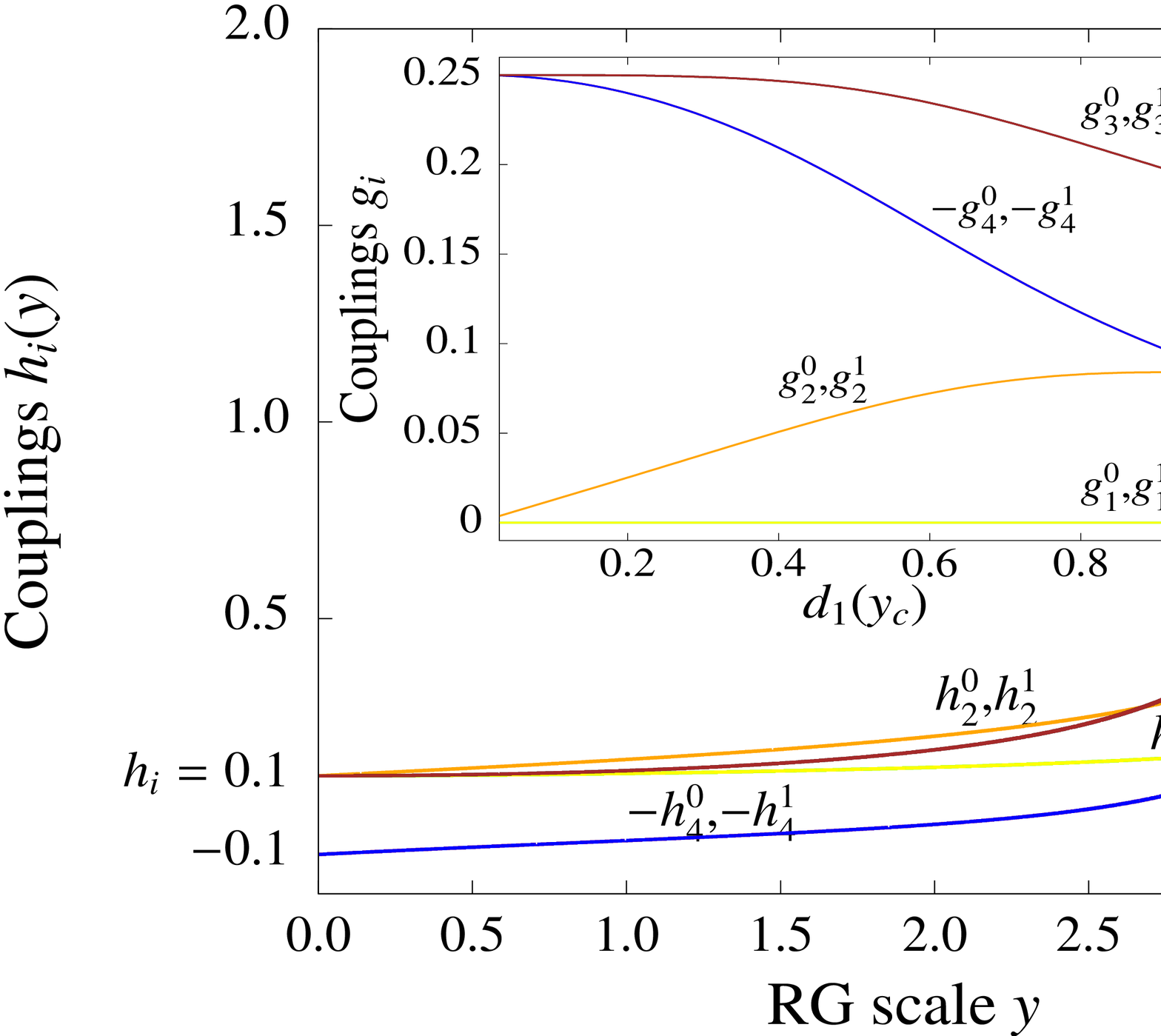}

\caption{\label{fig:all}Flow of couplings with renormalization group scale
$y$, starting with repulsive interactions, where the couplings in
different angular momentum channels($h_{r}^{0}$ and $h_{r}^{1}$)
are assumed to be degenerate initially, at $(h_{r}^{0,1})_{initial}=0.1$.
We find pair hopping between patches ($h_{3}$) and on-patch scattering
($h_{4}$) to be the dominant scattering channels. Here, the critical
point $y_{c}\approx3.65$.\protect \\
 The inset shows the evolution of the fixed-point couplings $g_{r}^{\ell}$($\ell=0,1$)
as a function of $d_{1}(y_{c})$($=\frac{1}{\sqrt{1+y_{c}}}$), which
is the ratio of the particle-hole to particle-particle susceptibilities
at the fixed point $y_{c}$. }
\end{figure}

To study the possible instabilities in this system, we construct a
two-patch renormalization group for the interaction vertices. In the
RG analysis, the instability is indicated in the form of a pole in
the vertex function. We consider only the electrons near the saddle
points at $\overline{X_{1}}$ and $\overline{X_{2}}$ on the (001)
surface. In our RG analysis, we distinguish between coupling constants
with different spin combinations ($h_{r}^{\sigma\sigma}$ and $h_{r}^{\sigma\overline{\sigma}}$,
or equivalently $h_{r}^{0}$ and $h_{r}^{1}$ respectively) and write
separate RG equations for the two kinds of interactions. 

We perform RG analysis up to one-loop level, integrating out high-energy
degrees of freedom gradually from an energy cutoff $\Lambda$, which
is the bandwidth. The susceptibilities in the different channels schematically
behave as $\chi_{0}^{pp}(\omega)\sim\ln[\Lambda/\omega]\ln[\Lambda/\text{max}(\omega,\mu)]$,
$\chi_{Q}^{ph}(\omega)\sim\ln[\Lambda/\text{max}(\omega,\mu)]\ln[\Lambda/\text{max}(\omega,\mu,t)]$
and $\chi_{0}^{ph}(\omega),\chi_{Q}^{pp}(\omega)\sim\ln[\Lambda/\text{max}(\omega,\mu)]$,
where $\omega$ denotes the energy away from the Van Hove singularities
and $t$ represents terms in the Hamiltonian that destroy the perfect
nesting. 

\begin{figure}
\includegraphics[width=0.9\columnwidth]{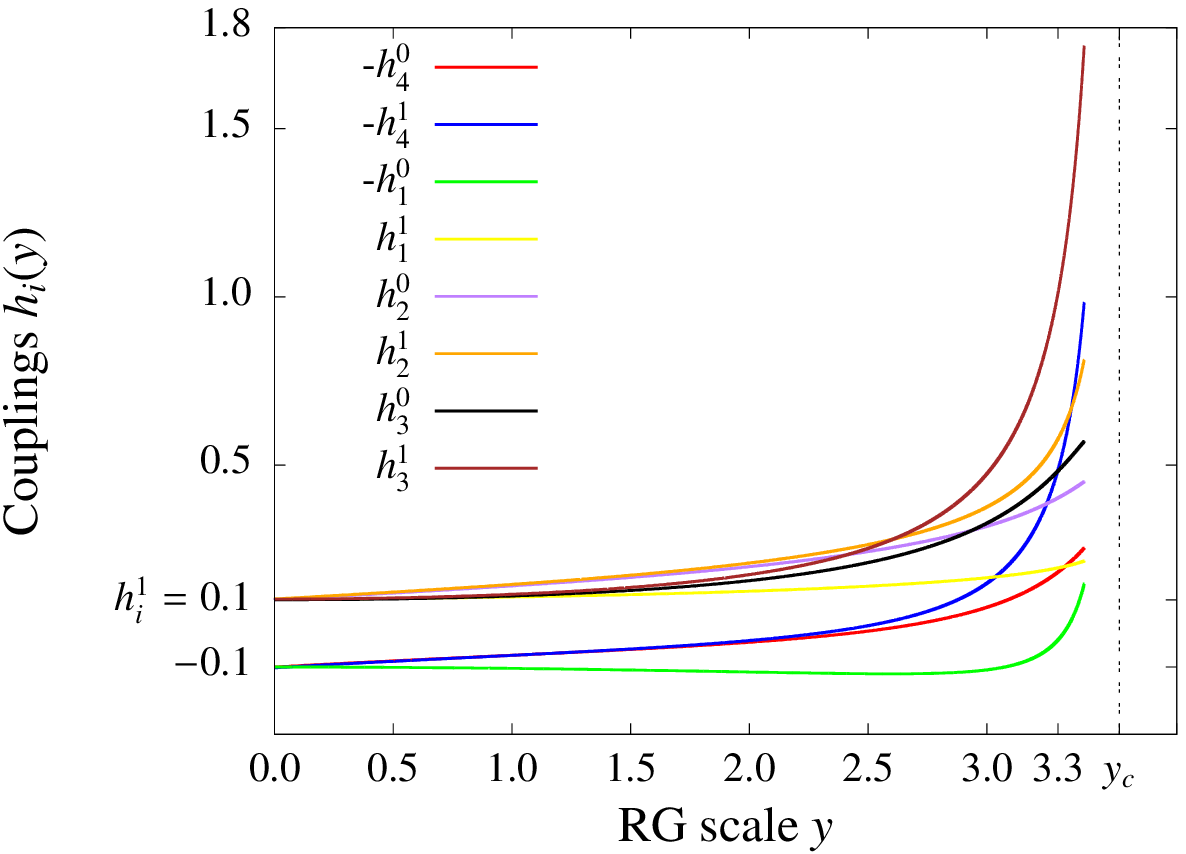}

\caption{\label{fig:g8}Flow of couplings with renormalization group scale
$y$, starting with repulsive interactions, where the $\ell=1$ components
of all the couplings are chosen to larger than the $\ell=0$ components
by $2\%$ initially, i.e. $\frac{|h_{r}^{1}-h_{r}^{0}|}{|h_{r}^{0}|}=0.02$,
where $(h_{r}^{0})_{initial}=0.1$. We find the $\ell=1$ components
of pair hopping between patches ($h_{3}$) and on-patch scattering
($h_{4}$) to be the most dominant couplings in this case. Here, the
critical point $y_{c}\approx3.56$. }
\end{figure}

We use $y\equiv\ln^{2}[\Lambda/\omega]\sim\chi_{0}^{pp}$ as the RG
flow parameter, and describe the relative weight of the other channels
as $d_{1}(y)=\frac{d\chi_{Q}^{ph}}{dy}$, $d_{2}(y)=\frac{d\chi_{0}^{ph}}{dy}$
and $d_{3}(y)=-\frac{d\chi_{Q}^{pp}}{dy}$, where $d_{1}(y)$ is taken
to be a function $\frac{1}{\sqrt{1+y}}$\cite{nandkishore2012chiral},
interpolating smoothly in between the limits $d_{1}(y=0)=1$ and $d_{1}(y\gg1)=\frac{1}{\sqrt{y}}$
, and $d_{2},d_{3}\ll d_{1}$. The multiplicative factor $d_{1}(y)$
essentially incorporates the effects of imperfect nesting in our analysis.
The RG equations are obtained by evaluating second-order diagrams
and collecting the respective combinatoric prefactors, for each of
the interactions $h_{1}$,$h_{2}$,$h_{3}$ and $h_{4}$. The diagrams
corresponding to the renormalization of the interaction $h_{2}$ are
shown in Fig. 7 in the Supplementary as an illustrative example. The
RG equations obtained are given by (where we have used the notation
$\sigma\sigma\equiv0$ and $\sigma\overline{\sigma}\equiv1$ for each
of the couplings)
\begin{align}
\frac{dh_{1}^{0}}{dy} & =2d_{1}(-(h_{1}^{0})^{2}-(h_{3}^{1})^{2}-(h_{1}^{1})^{2}\nonumber \\
 & +2h_{1}^{0}h_{2}^{0}+(h_{3}^{0})^{2}),\\
\frac{dh_{1}^{1}}{dy} & =2d_{1}(-2h_{1}^{0}h_{1}^{1}+2h_{1}^{1}h_{2}^{0}),\\
\frac{dh_{2}^{0}}{dy} & =2d_{1}((h_{2}^{0})^{2}+(h_{3}^{0})^{2}),\\
\frac{dh_{2}^{1}}{dy} & =2d_{1}((h_{2}^{1})^{2}+(h_{3}^{1})^{2}),
\end{align}
\begin{align}
\frac{dh_{3}^{0}}{dy} & =-4h_{4}^{0}h_{3}^{0}+2d_{1}(4h_{2}^{0}h_{3}^{0}\nonumber \\
 & -2h_{1}^{1}h_{3}^{1}),\\
\frac{dh_{3}^{1}}{dy} & =-4h_{4}^{1}h_{3}^{1}+2d_{1}(2h_{2}^{1}h_{3}^{1}\nonumber \\
 & -2h_{1}^{0}h_{3}^{1}+2h_{2}^{0}h_{3}^{1}),\\
\frac{dh_{4}^{0}}{dy} & =-2((h_{4}^{0})^{2}+(h_{3}^{0})^{2}),\\
\frac{dh_{4}^{1}}{dy} & =-2((h_{4}^{1})^{2}+(h_{3}^{1})^{2}).
\end{align}

\begin{figure}
\includegraphics[width=1.0\columnwidth]{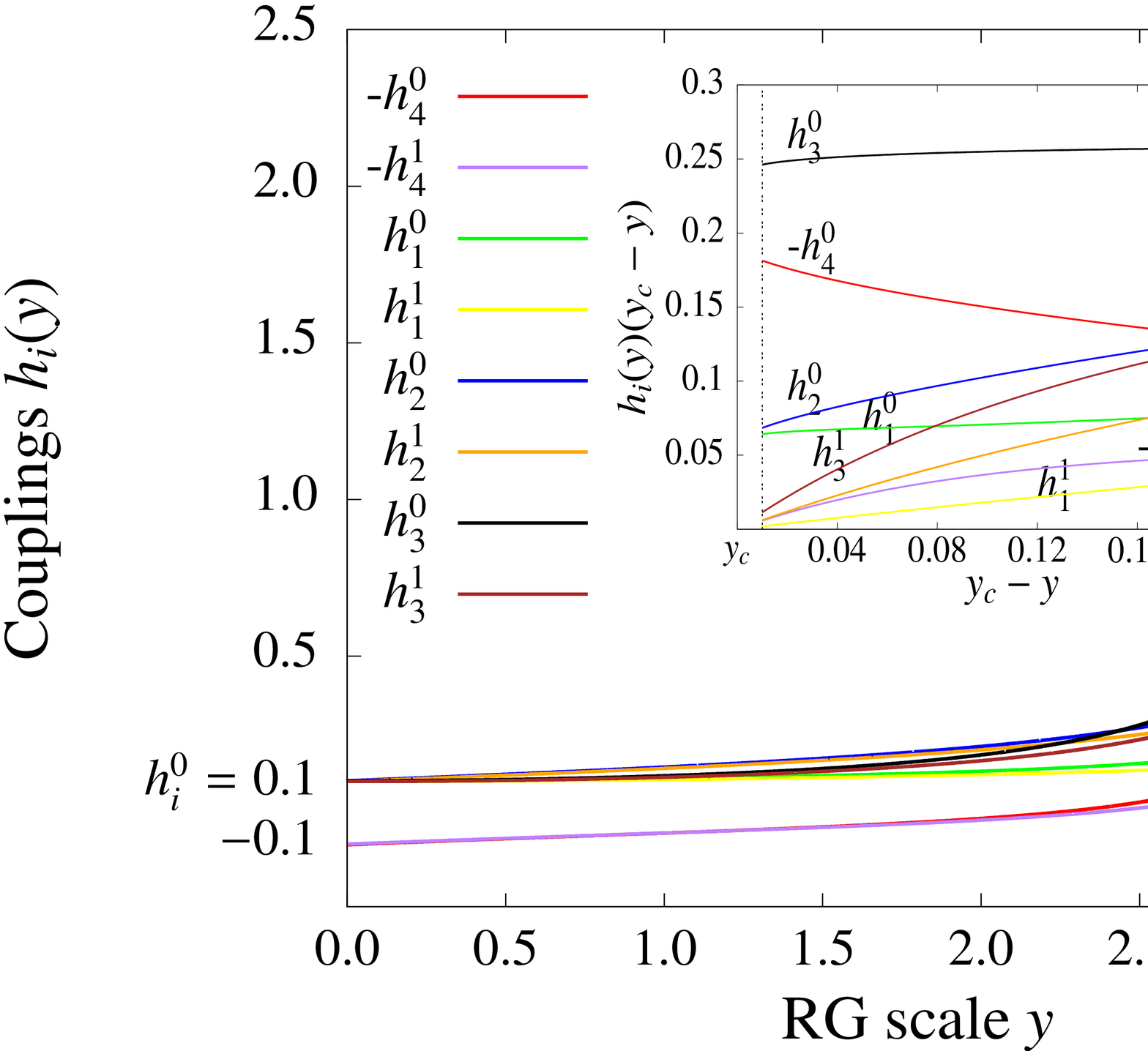}

\caption{\label{fig:g7}Flow of couplings with renormalization group scale
$y$, starting with repulsive interactions, where the $\ell=0$ components
of all the couplings are chosen to be larger than the $\ell=1$ components
by $2\%$ initially, i.e. $\frac{|h_{r}^{0}-h_{r}^{1}|}{|h_{r}^{1}|}=0.02$,
where $(h_{r}^{1})_{initial}=0.1$. We find the $\ell=0$ components
of pair hopping between patches ($h_{3}$) and on-patch scattering
($h_{4}$) to be the most dominant couplings in this case. Here, the
critical point $y_{c}\approx3.4$. \protect \\
The inset shows the behavior of $h_{r}(y)(y_{c}-y)$ as a function
of $(y_{c}-y)$ close to the fixed point $y_{c}$. The $y$-intercepts
of the different curves show the fixed-point values $g_{r}^{\ell}$
for the couplings $h_{r}^{\ell}(y)$. Evidently, the dominant couplings
near the critical point correspond to the $\ell=0$ channel in this
case.}
\end{figure}

These coupled differential equations are then solved, starting from
initial values of interactions in the weak-coupling regime($h_{r}^{0}=h_{r}^{1}\sim0.1$).
The results for the cases where (a) the couplings $h_{r}^{\ell}$
are degenerate for $\ell=0$ and $\ell=1$, (b) the couplings $h_{r}^{\ell}$
in the $\ell=1$ channel are chosen to dominate initially, (c) the
couplings $h_{r}^{\ell}$ in the $\ell=0$ channel are chosen to dominate
initially, are shown in the Figures \ref{fig:all},\ref{fig:g8} and
\ref{fig:g7} respectively.\textbf{ }The figures show results for
a Hund's splitting of $2\%$, and we have verified that even for a
splitting of $0.1\%$ introduced initially between the interactions
in the $\ell=0$ and $\ell=1$ channels ($\frac{|h^{1}-h^{0}|}{|h^{1}+h^{0}|}\sim0.1\%$)
the final set of dominant couplings $g_{r}^{\ell}$ near the critical
point of the RG correspond to the value of $\ell$ which has been
chosen to dominate initially. Thus, the results of our RG analysis
are found to be extremely sensitive to the sign of the Hund's splitting.
In contrast, the results are remarkably insensitive to the magnitude
as well as sign of an initial splitting introduced between the couplings
$h_{r}$ corresponding to the different scattering channels $r=1-4$.
This is graphically depicted in Fig.10 in the Supplementary.\\
 
\begin{figure}
\includegraphics[width=0.95\columnwidth]{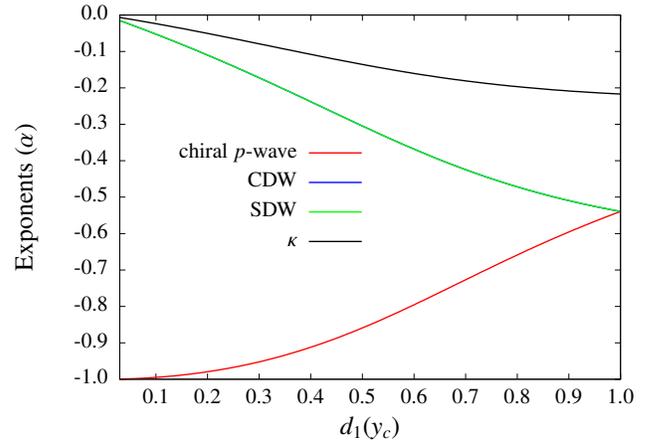}

\caption{\label{fig:exponents}The exponents $\alpha$, which are negative,
corresponding to the various susceptibilities: chiral $p$-wave superconductivity,
CDW, SDW and uniform charge compressibility ($\kappa$), plotted as
a function of $d_{1}(y_{c})$ for the case where each of the couplings
$g_{r}^{\ell}$ for $r=1-4$ and $\ell=0,1$ are degenerate. The order
of these exponents indicates that chiral $p$-wave superconductivity
is the leading instability (with the most negative exponent $\alpha_{pw}$)
throughout, and CDW and SDW have nearly the same values of exponents
$\alpha$ in this case.}
\end{figure}

We now investigate the instabilities of the system by evaluating the
susceptibilities $\chi$ for various types of order, introducing infinitesimal
test vertices corresponding to different kinds of pairing into the
action, such as $\triangle_{a}\psi_{a\sigma}^{\dagger}\psi_{a\sigma^{\prime}}^{\dagger}+\triangle_{a}^{*}\psi_{a\sigma}\psi_{a\sigma^{\prime}}$
for the patch $a=1,2$ (where the spin labels $\sigma,\sigma^{\prime}$
are meant to simply denote the presence or absence of the phase factors
$\exp[i\theta_{k}]$) corresponding to particle-particle pairing on
the patch \cite{nandkishore2012chiral}. 

The renormalization of the test vertex for particle-particle pairing
on a patch is governed by the equation \cite{nandkishore2012chiral}
\begin{equation}
\frac{\partial}{\partial y}\left(\begin{array}{c}
\Delta_{1}\\
\Delta_{2}
\end{array}\right)=2\left(\begin{array}{cc}
h_{4}^{1} & h_{3}^{1}\\
h_{3}^{1} & h_{4}^{1}
\end{array}\right)\left(\begin{array}{c}
\Delta_{1}\\
\Delta_{2}
\end{array}\right)\label{eq:13}
\end{equation}
 since we can only consider Cooper pairing in the $p$-wave channel
for spinless electrons. By transforming to the eigenvector basis,
we can obtain different possible order parameters, and choose the
one corresponding to the most negative eigenvalue. The vertices with
positive eigenvalues are suppressed under RG flow. 

At an electronic instability, the most divergent susceptibility $\chi$
determines the nature of the ordered phase. Each of the couplings
associated with the RG flow has an asymptotic form $h_{r}^{\ell}(y)=\frac{g_{r}^{\ell}}{y_{c}-y}$
near the instability threshold. The coefficients $g_{r}^{\ell}$ can
be determined as a function of $d_{1}(y_{c})$ (the results for the
case, where we start with identical initial values for each of the
couplings, are shown in the inset in Fig. \ref{fig:all}). We diagonalize
the Eq. \ref{eq:13} above and substitute the asymptodic form of the
interactions in the most negative eigenvalue. This gives us the exponent
$\alpha$ for the divergence of the susceptibility $\chi\propto(y_{c}-y)^{\alpha}$
for $p$-wave superconductivity. Likewise we can introduce test vertices
for other possible instabilities and obtain the corresponding exponents
for their susceptibilities \cite{seesi}. The exponents for intrapatch
$p$-wave pairing, charge-density wave, spin-density wave, uniform
spin, charge compressibility ($\kappa$) and finite-momentum $\pi$
pairing are given by-
\begin{align}
\alpha_{\text{pw}} & =2(-g_{3}^{1}+g_{4}^{1}),\nonumber \\
\alpha_{\text{CDW}} & =-2(g_{3}^{1}-g_{1}^{0}-g_{1}^{1}+g_{2}^{0})d_{1}(y_{c}),\nonumber \\
\alpha_{\text{SDW}} & =-2(g_{3}^{1}+g_{2}^{1})d_{1}(y_{c}),\nonumber \\
\alpha_{\kappa} & =-2(-g_{4}^{1}-(g_{1}^{0}-g_{2}^{0}-g_{2}^{1}))d_{2}(y_{c}),\nonumber \\
\alpha_{s} & =-2(g_{4}^{1}+g_{1}^{1})d_{2}(y_{c}),\nonumber \\
\alpha_{\pi}^{0} & =2(g_{2}^{0}-g_{1}^{0})d_{3}(y_{c}),\nonumber \\
\alpha_{\pi}^{1} & =2(g_{2}^{1}-g_{1}^{1})d_{3}(y_{c}).\label{eq:14}
\end{align}
 The $p$-wave order here is chiral since its symmetry is dictated
by the aforementioned $\exp[i\theta_{k}]$ dependence of the Berry
phase factors in the wave functions. It is important to note that
we have $p$-wave order on the patches, unlike \cite{PhysRevB.92.035132}
and \cite{PhysRevB.93.155108}. Consequently, this is a finite-momentum
pairing, with each patch $\overline{X_{i}}$ located at a finite momentum
with respect to the $\overline{\Gamma}$ point on the surface. Furthermore,
the relative phase of the $p$-wave order on different patches is
$\pi$, which means that we have $d$-wave order between the patches
\cite{seesi}. 

Figure \ref{fig:exponents} shows the behavior of the exponents for
$p$-wave pairing, SDW, CDW and charge compressibility as a function
of $d_{1}(y_{c})$. Comparison between the values of these exponents
shows that the most divergent susceptibility is $p$-wave superconductivity
throughout the parameter range $0<d_{1}(y_{c})<1$. The CDW and SDW
instabilities show a weaker divergence, and are followed by charge
compressibility. The exponents for uniform spin susceptibility and
$\pi$ pairing are always positive and hence, these orders are suppressed.
In the case of perfect nesting, i.e $d_{1}=1$, the SDW and CDW instabilities
become degenerate with $p$-wave superconductivity. 

Now, if a finite Hund's splitting is introduced initially such that
$h_{r}^{1}>h_{r}^{0}$, the above analysis holds and $p$-wave superconductivity
is still the dominant instability. However, for an initial Hund's
splitting of the opposite sign, i.e. $h_{r}^{0}>h_{r}^{1}$, we find
that the dominant couplings $g_{r}^{\ell}$ at the instability threshold
correspond to $\ell=0$. In this case, the exponents $\alpha$ for
each of the susceptibilities $\chi$ considered in Eq.\ref{eq:14}
turn out to be either positive or numerically close to zero. This
is due to subtle cancellations between contributions from the dominant
couplings in different scattering channels. Thus, none of the instabilities
considered above are found to occur in this case, within the one-loop
approximation. Clearly, the nature of instabilities in this system
is crucially dependent on the sign of the Hund's splitting. 

We now discuss the effects of weak disorder on superconductivity on
our crystalline topological insulator surface. Since potential scattering
of the electrons changes their momenta, we expect the $d$-wave pairing
across the patches to be sensitive to such disorder. However, within
a patch, the $p$-wave pairing is topologically protected. To see
this, note that our order parameter $<\psi_{k}\psi_{-k}>\sim\Delta_{0}\exp[i\theta_{k}]$
( where $\psi$ denotes the spinless fermion in the relevant band
and $\theta_{k}$ arises from the nontrivial Berry phases). Translated
to the valley-spin picture, this shows that the superconducting order
parameter in terms of those fermions has no momentum dependence, and
hence, cannot be degraded by weak potential disorder. The $p$-wave superconductivity is also found to survive in the presence of magnetic
impurities for a finite Hund's splitting of interactions \cite{Kundu2017Competing}.

Finally, we discuss the experimental implications of our work. Recently,
there have been reports of surface superconductivity induced on the
surface of Pb$_{0.6}$Sn$_{0.4}$Te by forming a mesoscopic point
contact using a nonsuperconducting metal \cite{das2016unexpected}.
The observed transition temperature is in the range 3.7-6.5 K. We
expect transition temperatures roughly an order of magnitude smaller
than the bandwidth $\Lambda$, which is of the order of the band gap.
However, the nature of the Cooper pair order in the experiment is
not yet settled and further experimental work needs to be done in
this direction to confirm our prediction of surface $p$-wave superconductivity
in this material. Recently, we have come across a paper \cite{Mazur2017Majorana}
which reports the detection of an electron-hole gap with a broad zero-bias
conductance maximum at the topological surfaces of diamagnetic, paramagnetic,
and ferromagnetic Pb$_{1-y-x}$Sn$_{y}$Mn$_{x}$Te (where $y\gtrsim0.67$
and $0\leq x<0.1$) using soft-contact spectroscopy. The MBS-like
conductance spectra obtained with and without magnetic impurities
are found to be intrinsic in origin, which we believe supports our claim. Our approach could also be useful for studying phase
competition in other two-dimensional systems with multiple Fermi patches
in the presence of Hund's splitting. In particular, this could be
relevant for Type-II Dirac surface states on certain surfaces of antiperovskites\cite{PhysRevB.95.035151},
or for the bulk band structure of the Dirac semimetal Na$_{3}$Bi
with multiple Dirac nodes connecting via a Lifshitz point\cite{PhysRevB.92.075115},
in a quasi-2D approximation.
\begin{acknowledgments}
The authors gratefully acknowledge useful discussions with Kedar Damle
and Rajdeep Sensarma. SK acknowledges Debjyoti Burdhan for his help
with some of the figures. VT acknowledges DST for a Swarnajayanti
grant (No. DST/SJF/PSA-0212012-13). 
\end{acknowledgments}

\appendix
\begin{widetext}
\textbf{Supplementary material for Role of Hund's splitting in electronic
phase competition in Pb$_{1-x}$Sn$_{x}$Te:}

Here we provide additional information on 1) electron interactions
in the valley-spin picture and effective interactions when projected
to a band and 2) RG equations for test vertices corresponding to different
kinds of pairing, and 3) Fixed point values of different couplings
as a function of $d_{1}(y_{c})$

\textbf{Interactions between electrons in the valley-spin basis:}

Here we derive the effective interaction model obtained upon projecting
the interactions in the valley-spin basis to one of the surface bands
(the positive energy band closest to the saddle points) for each of
the $\overline{X}$ points. The interaction Hamiltonian for surface
electrons with valley and spin labels is given by \textbf{
\begin{equation}
H_{I}=\frac{1}{2}\sum_{a,b,c,d,\sigma,\sigma^{\prime}}U_{abcd}^{\sigma\sigma^{\prime}}c_{\sigma a}^{\dagger}c_{\sigma^{\prime}b}^{\dagger}c_{\sigma^{\prime}c}c_{\sigma d}\label{eq:1}
\end{equation}
}where $a,b,c,d$ refer to different valleys (which are either all
the same, same in pairs or all different in the above sum) and $\sigma,\sigma^{\prime}$
refer to spins. Here, we consider $U_{abcd}^{\sigma\sigma^{\prime}}=U_{1}^{\sigma\sigma^{\prime}}$when
$(a,c)$ belong to one $\overline{X}$-point (i.e. the L-valleys corresponding
to $(a,c)$ are projected to one of the $\overline{X}$-points) and
$(b,d)$ belong to the other $\overline{X}$-point. Similarly, $U_{abcd}^{\sigma\sigma^{\prime}}=U_{2}^{\sigma\sigma^{\prime}}$
when $(b,c)$ belong to one $\overline{X}$-point and $(a,d)$ belong
to the other, $U_{3}^{\sigma\sigma^{\prime}}$when $(a,b)$ belong
to one $\overline{X}$-point and $(c,d)$ to the other, and $U_{4}^{\sigma\sigma^{\prime}}$when
$a$,$b$,$c$ and $d$ all correspond to L-points projected to the
same $\overline{X}$-point. The interactions depend only on the relative
orientations of the spins, for example, $U^{\sigma\sigma^{\prime}}$
can be written as $U^{\sigma\sigma}\delta_{\sigma\sigma^{\prime}}+U^{\sigma\overline{\sigma}}(1-\delta_{\sigma\sigma^{\prime}})$.
For the k.p Hamiltonian $H_{\overline{X_{1}}}(k)$ and $H_{\overline{X_{2}}}(k)$
of the (001) surface, the operators corresponding to different bands
can be rewritten in terms of the operators for different valley and
spin combinations as follows
\begin{align}
\psi_{1} & =A_{1}c_{\uparrow1}+B_{1}c_{\downarrow1}+C_{1}c_{\uparrow2}+D_{1}c_{\downarrow2},\nonumber \\
\psi_{2} & =A_{2}c_{\uparrow1}+B_{2}c_{\downarrow1}+C_{2}c_{\uparrow2}+D_{2}c_{\downarrow2},\nonumber \\
\psi_{3} & =A_{3}c_{\uparrow1}+B_{3}c_{\downarrow1}+C_{3}c_{\uparrow2}+D_{3}c_{\downarrow2},\nonumber \\
\psi_{4} & =A_{4}c_{\uparrow1}+B_{4}c_{\downarrow1}+C_{4}c_{\uparrow2}+D_{4}c_{\downarrow2},\nonumber \\
\psi_{5} & =A_{5}c_{\uparrow3}+B_{5}c_{\downarrow3}+C_{5}c_{\uparrow4}+D_{5}c_{\downarrow4},\nonumber \\
\psi_{6} & =A_{6}c_{\uparrow3}+B_{6}c_{\downarrow3}+C_{6}c_{\uparrow4}+D_{6}c_{\downarrow4},\nonumber \\
\psi_{7} & =A_{7}c_{\uparrow3}+B_{7}c_{\downarrow3}+C_{7}c_{\uparrow4}+D_{7}c_{\downarrow4},\nonumber \\
\psi_{8} & =A_{8}c_{\uparrow3}+B_{8}c_{\downarrow3}+C_{8}c_{\uparrow4}+D_{8}c_{\downarrow4},\label{eq:2}
\end{align}
where $\{A_{i},B_{i},C_{i},D_{i},i=1$ to $8\}$ correspond to the
complex conjugates of the nonzero components of the different normalized
energy eigenvectors, and are functions of $k_{x}$ and $k_{y}$ in
the two-dimensional momentum space. We denote the L-valleys projected
to one of the $\overline{X}$-points by $1$ and $2$, and those projected
to the other point by $3$ and $4$. Thus, the total number of bands
is eight. Since the points $\overline{X}$ are decoupled from each
other, four of these components for each eigenvector vanish, giving
rise to the expression in Eq. \ref{eq:2}. We can invert the above
equations to write the $c_{\alpha a}'s$ in terms of $\psi_{i}'s$.
Substituting all of these expressions into $H_{I}$ in Eq. \ref{eq:1}
above, and writing $c_{\alpha a}$ as $\sum_{i}u_{\alpha ai}\psi_{i}$,
we have
\begin{align}
H_{I} & =\frac{1}{2}(\sum_{a,b,c,d,\sigma,\sigma^{\prime}}\sum_{i,j,k,l}U_{abcd}^{\sigma\sigma^{\prime}}u_{\sigma ai}^{*}(k_{1}^{\prime})u_{\sigma^{\prime}bj}^{*}(k_{2}^{\prime})\nonumber \\
 & \qquad\times u_{\sigma^{\prime}ck}(k_{1})u_{\sigma dl}(k_{2})\psi_{i}^{\dagger}\psi_{j}^{\dagger}\psi_{k}\psi_{l})\label{eq:3-1}
\end{align}

where $k_{1},k_{2},k_{1}^{\prime},k_{2}^{\prime}$ are constrained
by momentum conservation, and $i$,$j$,$k$ and $l$ refer to the
various bands, and $(a,b,c,d)$ are either all the same, same in pairs
or all different in the above sum. Now, we are only interested in
the two bands (for a given $\overline{X}$-point) which lie in the
bulk band gap and are closer to the saddle points in energy. In particular,
we shall concentrate on the positive energy bands lying closer to
the saddle points for each of the $\overline{X}$ points, in which
case we can drop all the terms from the above equations except those
involving $\psi_{2}$ and $\psi_{6}$, the relevant bands in our case.
We then have $c_{\uparrow1}=u_{\uparrow1}\psi_{2}$, $c_{\downarrow1}=u_{\downarrow1}\psi_{2}$,
$c_{\uparrow2}=u_{\uparrow2}\psi_{2}$ and $c_{\downarrow2}=u_{\downarrow2}\psi_{2}$,
and likewise for $\psi_{6}$ with the valleys $3$ and $4$, suppressing
the contributions from the other bands. Considering only the contributions
from the two lower positive energy bands (corresponding to the two
$\overline{X}$ points) which are degenerate, the above Eq. \ref{eq:3-1}
can be rewritten as
\begin{align}
H_{I} & =\sum_{i}\sum_{\sigma,\sigma^{\prime}}\frac{1}{2}h_{4}^{\sigma\sigma^{\prime}}\psi_{i}^{\dagger}\psi_{i}^{\dagger}\psi_{i}\psi_{i}\nonumber \\
 & \qquad+\sum_{i\neq j}\sum_{\sigma,\sigma^{\prime}}\frac{1}{2}(h_{1}^{\sigma\sigma^{\prime}}\psi_{i}^{\dagger}\psi_{j}^{\dagger}\psi_{i}\psi_{j}\nonumber \\
 & \qquad+h_{2}^{\sigma\sigma^{\prime}}\psi_{i}^{\dagger}\psi_{j}^{\dagger}\psi_{j}\psi_{i}+h_{3}^{\sigma\sigma^{\prime}}\psi_{i}^{\dagger}\psi_{i}^{\dagger}\psi_{j}\psi_{j})\label{eq:4}
\end{align}
where the sum is over the two low-energy bands only, and $h_{1}^{\sigma\sigma^{\prime}}=\sum_{a,b,c,d}U_{1}^{\sigma\sigma^{\prime}}u_{\sigma ai}^{*}(k_{1}^{\prime})u_{\sigma^{\prime}bj}^{*}(k_{2}^{\prime})u_{\sigma^{\prime}ci}(k_{1})u_{\sigma dj}(k_{2})$
(where one of the low-energy bands denoted by $i$ has nonzero components
for valleys $(a,c)$ and the other one denoted by $j$ for valleys
$(b,d)$), and this gives us the corresponding coupling $h_{1}$ used
in the low-energy theory in Eq.(3) of the main text, when scaled with
respect to the number of such combinations of valleys. The rest of
the couplings $h_{2}$, $h_{3}$ and $h_{4}$ can be similarly defined
in terms of the interactions in the valley-spin picture and the form
factors for the basis transformation. Thus, there are four kinds of
allowed scattering terms between electrons belonging to the two bands
under consideration. These correspond to exchange processes between
electrons on the two different bands($h_{1}$), Coulomb interaction
between electrons on different bands ($h_{2}$), pair hopping between
the two bands ($h_{3}$) and scattering between different valleys
within a band ($h_{4}$).

\begin{figure}
\begin{centering}
\includegraphics[width=0.5\columnwidth]{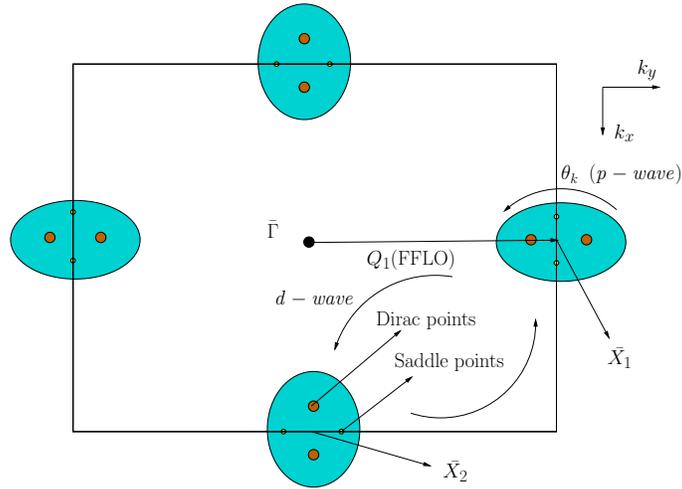}
\par\end{centering}

\caption{\label{fig:schematic}A schematic representation of the (001) surface
with two Dirac points and saddle points each at the $\overline{X_{1}}$and
$\overline{X_{2}}$ points. The FFLO wave vector connecting the $\overline{X_{1}}$
point to the origin $\overline{\Gamma}$ is represented by $Q_{1}$.
In our analysis, we obtain a phase factor of $\exp[i(\theta_{k}-\theta_{k^{\prime}})]$,with
respect to the $\overline{X}$ points, associated with the effective
interactions in the band picture, while the relative phases of the
order parameter between the patches $\overline{X_{1}}$ and $\overline{X_{2}}$
is $\pi$ ($d$-wave)(with respect to the $\overline{\Gamma}$ point).
The superconducting order parameter has the form $\Delta_{k}=\Delta_{0}\left(\protect\begin{array}{c}
1\protect\\
-1
\protect\end{array}\right)_{\overline{X}}\otimes\exp[i\theta_{k}]$. We have considered a situation where the Fermi surface of a patch
encloses both the Van-Hove points. Electrons anywhere in the patch
experience an enhanced density of states due to the proximity of one
or more Van-Hove points. }
\end{figure}

\begin{figure}
\includegraphics[width=0.8\columnwidth]{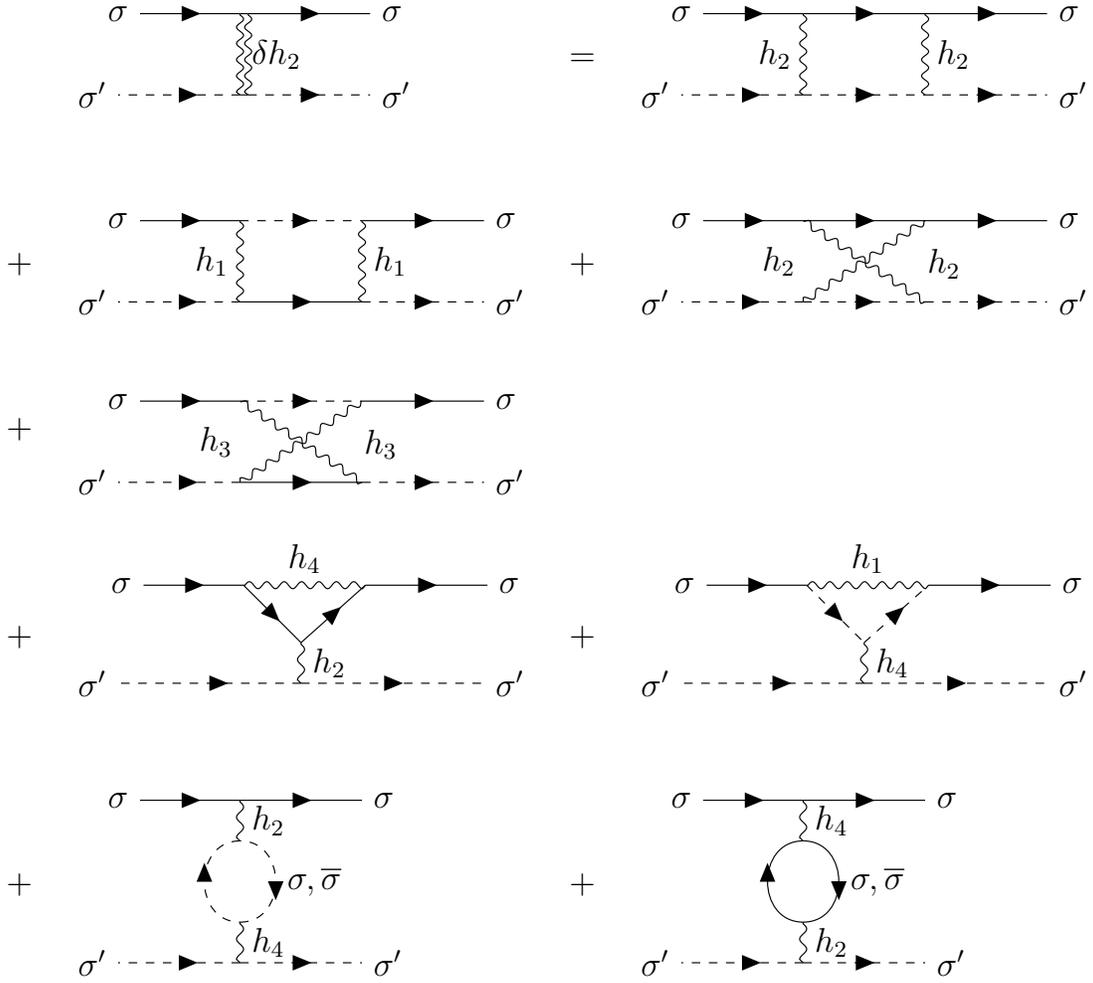}

\caption{\label{fig:h4rg}Diagrams for one-loop renormalization of the coupling
$h_{2}$. The diagrams for $h_{1}$, $h_{3}$ and $h_{4}$ are similarly
obtained.}
\end{figure}

\begin{figure}
\includegraphics[width=0.8\columnwidth]{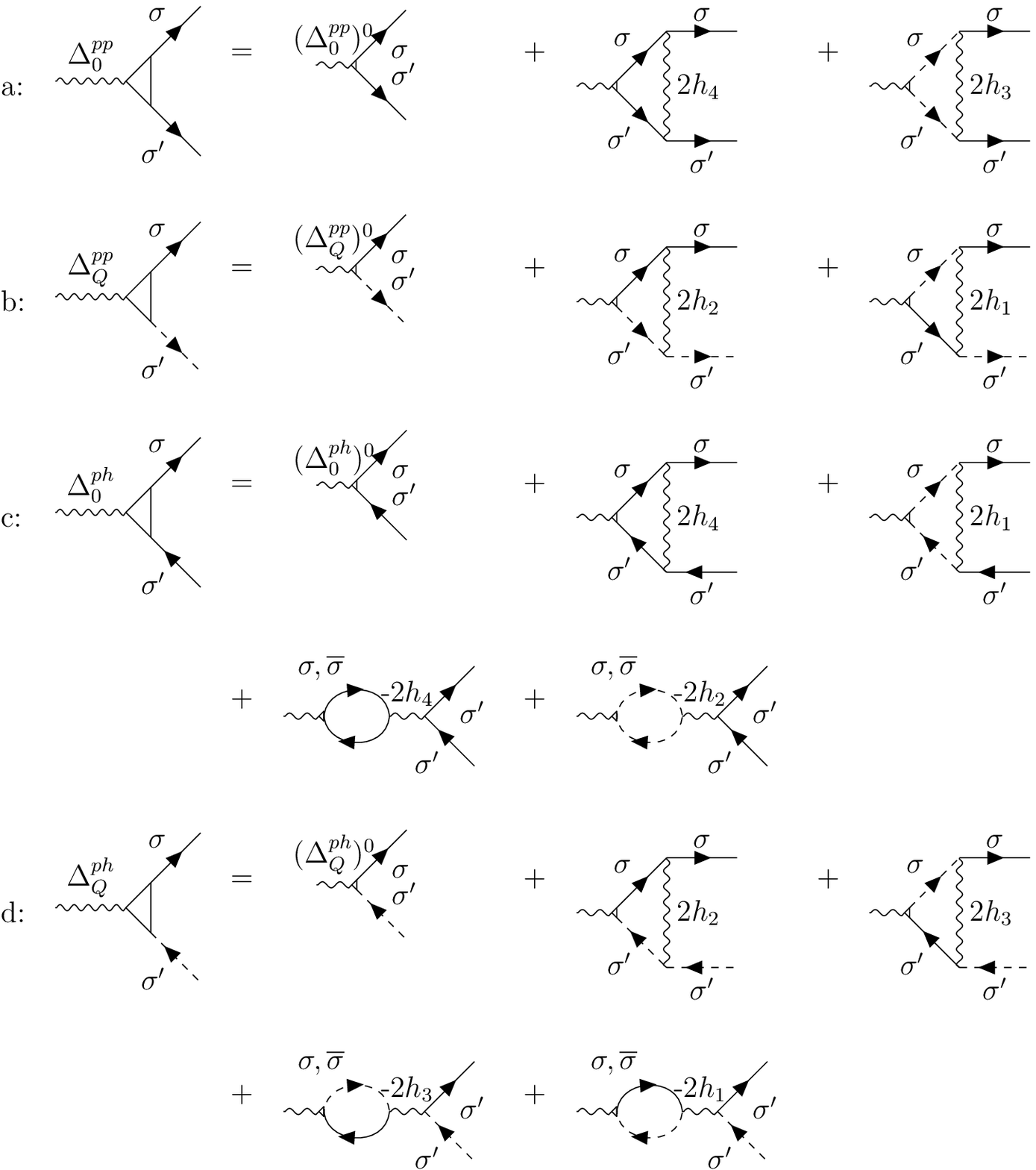}

\caption{\label{fig:Vertex-renormalization}Test vertex renormalization corresponding
to (a) particle-particle pairing on the patch, (b) particle-particle
pairing between patches, (c) particle-hole pairing on the patch and
(d) particle-hole pairing between patches, where $Q$ refers to the
nesting vector between the patches $\overline{X_{1}}$and $\overline{X_{2}}$
in two dimensions. }
\end{figure}

\textbf{Susceptibilities:}

The renormalization equations for the different kinds of ordering
considered, in the particle-particle as well as particle-hole channel,
are given as follows. 

The renormalization of the test vertex corresponding to particle-hole
pairing between the patches, in the $\ell=0$ channel is given by
\[
\frac{\partial}{\partial y}\left(\begin{array}{c}
\Delta_{12}\\
\Delta_{21}
\end{array}\right)=
\]
\begin{equation}
-2d_{1}(y)\left(\begin{array}{cc}
h_{2}^{0}-h_{1}^{0}-h_{1}^{1} & -h_{3}^{1}\\
-h_{3}^{1} & h_{2}^{0}-h_{1}^{0}-h_{1}^{1}
\end{array}\right)\left(\begin{array}{c}
\Delta_{12}\\
\Delta_{21}
\end{array}\right)\label{eq:5}
\end{equation}
and in the $\ell=1$ channel, by 
\begin{equation}
\frac{\partial}{\partial y}\left(\begin{array}{c}
\Delta_{12}\\
\Delta_{21}
\end{array}\right)=-2d_{1}(y)\left(\begin{array}{cc}
h_{2}^{1} & h_{3}^{1}\\
h_{3}^{1} & h_{2}^{1}
\end{array}\right)\left(\begin{array}{c}
\Delta_{12}\\
\Delta_{21}
\end{array}\right)\label{eq:6}
\end{equation}
The renormalization of the test vertex corresponding to particle-particle
pairing between the patches, in the $\ell=0$ channel, is given by
\begin{equation}
\frac{\partial}{\partial y}\left(\begin{array}{c}
\Delta_{12}\\
\Delta_{21}
\end{array}\right)=2d_{3}(y)\left(\begin{array}{cc}
h_{2}^{0} & h_{1}^{0}\\
h_{1}^{0} & h_{2}^{0}
\end{array}\right)\left(\begin{array}{c}
\Delta_{12}\\
\Delta_{21}
\end{array}\right)\label{eq:7}
\end{equation}
and in the $\ell=1$ channel, by 
\begin{equation}
\frac{\partial}{\partial y}\left(\begin{array}{c}
\Delta_{12}\\
\Delta_{21}
\end{array}\right)=2d_{3}(y)\left(\begin{array}{cc}
h_{2}^{1} & h_{1}^{1}\\
h_{1}^{1} & h_{2}^{1}
\end{array}\right)\left(\begin{array}{c}
\Delta_{12}\\
\Delta_{21}
\end{array}\right)\label{eq:8}
\end{equation}
The renormalization of the test vertex corresponding to particle-hole
pairing on a patch, in the $\ell=0$ channel, is given by 
\[
\frac{\partial}{\partial y}\left(\begin{array}{c}
\Delta_{1}\\
\Delta_{2}
\end{array}\right)=
\]
\begin{equation}
-2d_{2}(y)\left(\begin{array}{cc}
-h_{4}^{1} & h_{1}^{0}-h_{2}^{0}-h_{2}^{1}\\
h_{1}^{0}-h_{2}^{0}-h_{2}^{1} & -h_{4}^{1}
\end{array}\right)\left(\begin{array}{c}
\Delta_{1}\\
\Delta_{2}
\end{array}\right)\label{eq:9}
\end{equation}
 and in the $\ell=1$ channel, is given by 
\begin{equation}
\frac{\partial}{\partial y}\left(\begin{array}{c}
\Delta_{1}\\
\Delta_{2}
\end{array}\right)=-2d_{2}(y)\left(\begin{array}{cc}
h_{4}^{1} & h_{1}^{1}\\
h_{1}^{1} & h_{4}^{1}
\end{array}\right)\left(\begin{array}{c}
\Delta_{1}\\
\Delta_{2}
\end{array}\right)\label{eq:10}
\end{equation}

The diagrams corresponding to the renormalization of the different
kinds of pairing vertices are shown in Fig. \ref{fig:Vertex-renormalization}.
The most negative eigenvalue for Cooper pairing on the patch is given
by $2(-h_{3}^{1}+h_{4}^{1})$ which corresponds to the eigenvector
$\frac{1}{\sqrt{2}}\left(\begin{array}{cc}
-1 & 1\end{array}\right)$, competing with those for CDW and SDW order, given by $-2(h_{3}^{1}-h_{1}^{0}-h_{1}^{1}+h_{2}^{0})d_{1}(y)$
(corresponding to the eigenvector $\frac{1}{\sqrt{2}}\left(\begin{array}{cc}
-1 & 1\end{array}\right)$) and $-2(h_{3}^{1}+h_{2}^{1})d_{1}(y)$ (corresponding to the eigenvector
$\frac{1}{\sqrt{2}}\left(\begin{array}{cc}
1 & 1\end{array}\right)$) respectively. This is followed by particle-hole pairing on a patch
in the $\ell$=0 channel, with the more negative eigenvalue given
by $-2(-h_{4}^{1}-(h_{1}^{0}-h_{2}^{0}-h_{2}^{1}))d_{2}(y)$ (corresponding
to the eigenvector $\frac{1}{\sqrt{2}}\left(\begin{array}{cc}
-1 & 1\end{array}\right)$ ). Thus, the dominant instability of our system, namely $p$-wave
superconductivity, appears in the $\ell=1$ channel. 

\begin{figure}
\includegraphics[width=0.45\columnwidth]{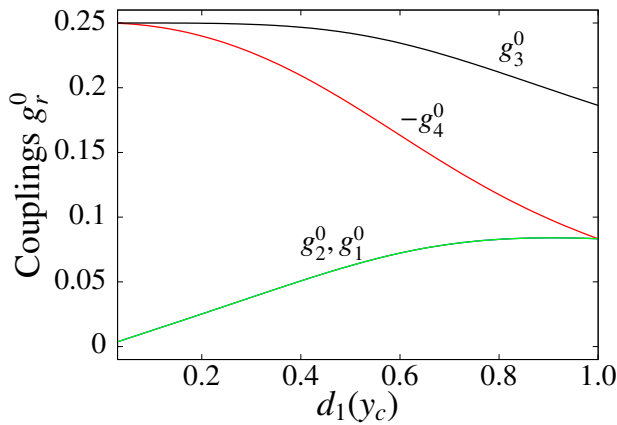}

\caption{\label{fig:gi0l=00003D1}The fixed point values for $g_{r}^{0}$ as
a function of $d_{1}(y_{c})$ for the case where the $\ell$=0 components
of all the couplings dominate initially. Note that the fixed point
values $g_{2}^{0}$ and $g_{1}^{0}$ turn out to be identical. }
\end{figure}

\begin{figure}
(a)\includegraphics[width=0.45\columnwidth]{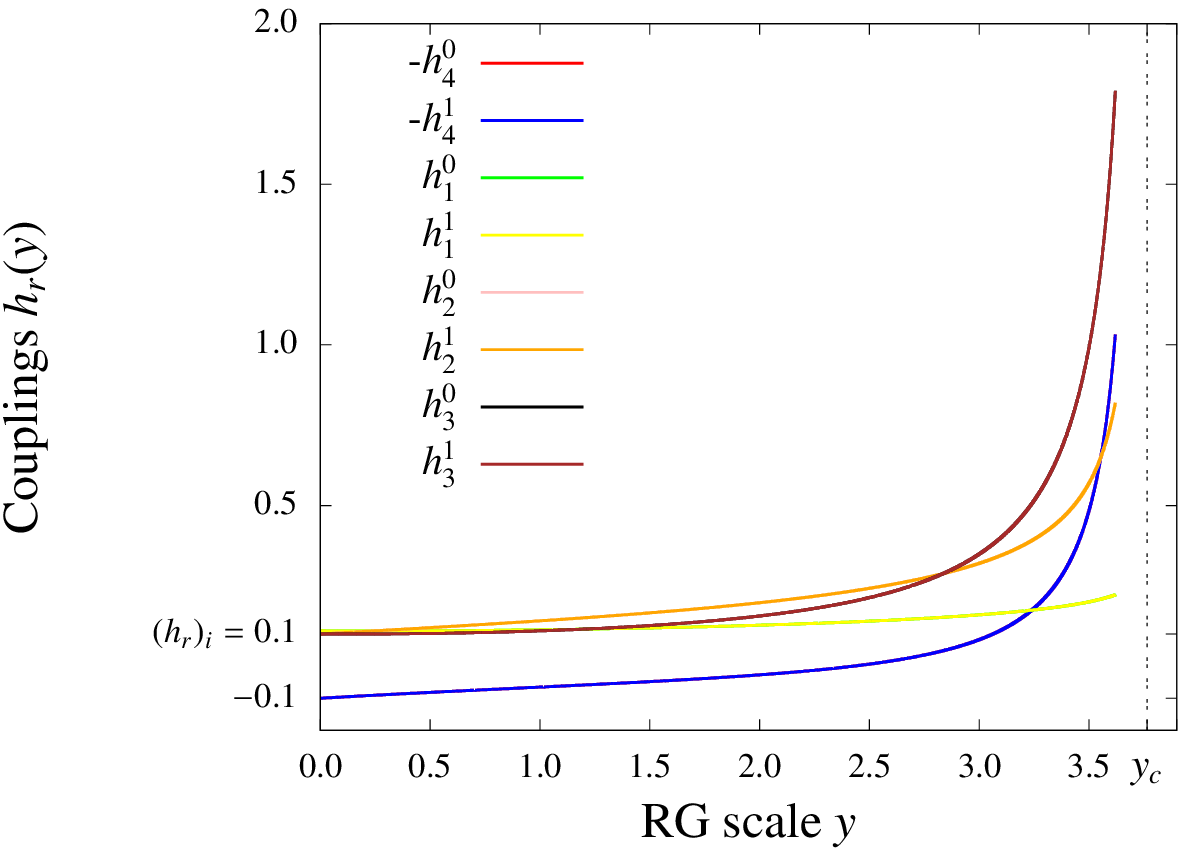}

\bigskip{}

(b)\includegraphics[width=0.45\columnwidth]{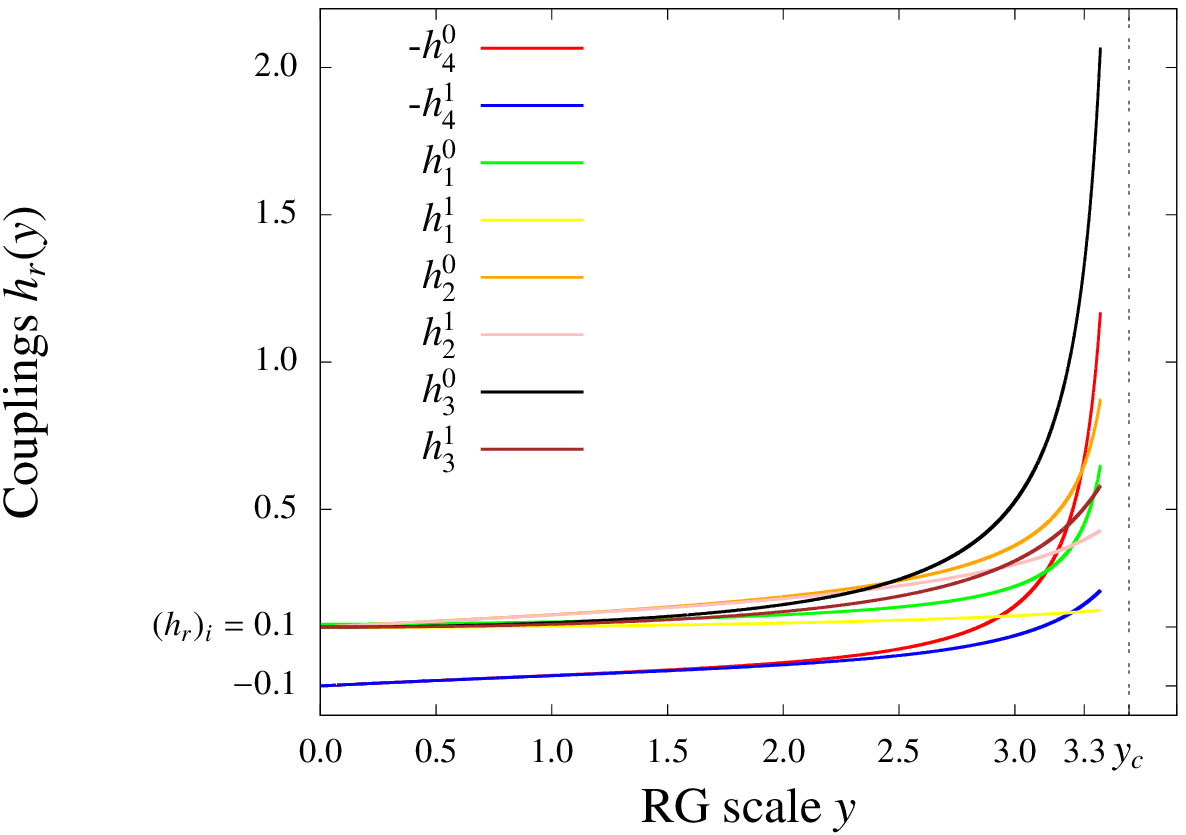}

\caption{\label{fig:g3g4}(a)Flow of the couplings with the RG scale $y$,
with an initial splitting in the different scattering channels $r$.
Here we have chosen the initial value of $h_{1}$ to be greater than
all the other $h_{r}$ by $10\%$, i.e. $\frac{|h_{1}^{\ell}-h_{r}^{\ell}|}{|h_{r}^{\ell}|}=0.1$($r\protect\neq1$)
for $\ell=0,1$, where $(h_{r}^{\ell})_{initial}=0.1$ for $r\protect\neq1$.
The resulting order of the couplings at the fixed-point $y_{c}$ is
identical to the case where all the couplings are chosen to be degenerate
initially (see Fig.2 in main text). This illustrates that our RG flows
are insensitive to the initial order of the couplings in different
scattering channels $r=1-4$ , as long as $h_{r}^{0}=h_{r}^{1}$ for
all $r$. Here, the critical point $y_{c}\approx3.8$.\protect \\
(b)Flow of the couplings with the RG scale $y$, with $h_{1}^{0}>h_{1}^{1}$
by $10\%$ initially, i.e. $\frac{|h_{1}^{0}-h_{1}^{1}|}{|h_{1}^{1}|}=0.1$,
where $(h_{r}^{0})_{initial}=0.1$ for $r\protect\neq1$ and $(h_{r}^{1})_{initial}=0.1$
for all $r$. This changes the order of the couplings at the fixed
point drastically, and the couplings $h_{3}^{0}$ and $(-h_{4}^{0})$
now dominate near the fixed point of RG flow. Here, the critical point
$y_{c}\approx3.5$. }
\end{figure}

\textbf{Fixed point values of couplings as a function of $d_{1}(y_{c})$:}

As discussed in the main text, the different couplings $h_{r}^{\ell}(y)$
have an asymptotic form $\frac{g_{r}^{\ell}}{y_{c}-y}$ near the critical
point $y_{c}$ of the RG flow. In order to determine the behavior
of the fixed point values $g_{r}^{\ell}$ for the different couplings
as a function of $d_{1}(y_{c})$, we substitute this asymptotic form
into the RG equations (Eq.4-11 of the main text) to obtain the polynomial
equations

\begin{align}
g_{1}^{0} & =2d_{1}(y_{c})(-(g_{1}^{0})^{2}-(g_{3}^{1})^{2}-(g_{1}^{1})^{2}\nonumber \\
 & +2g_{1}^{0}g_{2}^{0}+(g_{3}^{0})^{2}),\nonumber \\
g_{1}^{1} & =2d_{1}(y_{c})(-2g_{1}^{0}g_{1}^{1}+2g_{1}^{1}g_{2}^{0}),\nonumber \\
g_{2}^{0} & =2d_{1}(y_{c})((g_{2}^{0})^{2}+(g_{3}^{0})^{2}),\nonumber \\
g_{2}^{1} & =2d_{1}(y_{c})((g_{2}^{1})^{2}+(g_{3}^{1})^{2}),\nonumber \\
g_{3}^{0} & =-4g_{4}^{0}g_{3}^{0}+2d_{1}(y_{c})(4g_{2}^{0}g_{3}^{0}-2g_{1}^{1}g_{3}^{1}),\nonumber \\
g_{3}^{1} & =-4g_{4}^{1}g_{3}^{1}+2d_{1}(y_{c})(2g_{2}^{1}g_{3}^{1}\nonumber \\
 & -2g_{1}^{0}g_{3}^{1}+2g_{2}^{0}g_{3}^{1}),\nonumber \\
g_{4}^{0} & =-2(g_{4}^{0})-2(g_{3}^{0})^{2},\nonumber \\
g_{4}^{1} & =-2(g_{4}^{1})^{2}-2(g_{3}^{1})^{2}.\label{eq:11}
\end{align}

These coupled equations are then solved with appropriate initial conditions,
to determine $g_{r}^{\ell}$($\ell=0,1$) as a function of $d_{1}(y_{c})$,
which is the ratio of the particle-hole and particle-particle susceptibilities
at the fixed point $y_{c}$. The behaviour of $g_{r}^{\ell}$ as a
function of $d_{1}(y_{c})$ when all the couplings are chosen to be
degenerate initially, is shown in the inset in Fig.2 of the main text.
The corresponding behavior when the degeneracy between the couplings
in the $\ell=0$ and $\ell=1$ channels is lifted (such that $g_{r}^{0}>g_{r}^{1}$
for all $r$) is shown in Fig. \ref{fig:gi0l=00003D1} (here we have
only shown the behavior of the couplings $g_{r}^{0}$, as the fixed-point
values $g_{r}^{1}$ turn out to be very small in this case).

\end{widetext}

\bibliographystyle{apsrev4-1}

\end{document}